\documentclass[12pt]{article}
\usepackage[super,compress]{cite}
\usepackage[super]{cite}
\usepackage{amsmath,amsfonts,euscript,amssymb}
\usepackage{epsfig}

\DeclareSymbolFont{bfitletters}{OML}{cmm}{bx}{it}
\DeclareSymbolFont{bfitoperators}   {OT1}{cmr} {m}{n}
\DeclareMathSymbol{\bfitomega}{\mathord}{bfitletters}{"21}
\DeclareMathSymbol{\bfitrho}{\mathord}{bfitletters}{"1A}
\DeclareMathSymbol{\bfitgamma}{\mathord}{bfitletters}{"0D}
\DeclareMathSymbol{\bfitchi}{\mathord}{bfitletters}{"1F}
\DeclareMathSymbol{\bfitxi}{\mathord}{bfitletters}{"18}
\DeclareMathSymbol{\bfitpi}{\mathord}{letters}{"19}

\newcommand{\be}{\begin{equation}}
\newcommand{\ee}{\end{equation}}
\newcommand{\bea}{\begin{eqnarray}}
\newcommand{\eea}{\end{eqnarray}}

\begin{document}

\begin{center}
{\Large\bf EoS of Casimir vacuum of massive fields in Friedmann Universe}\\
\vskip 1cm
{\bf Alexander E. Pavlov}
\vskip 0.5cm
{\it Institute of Mechanics and Power Engineering,
  Russian State Agrarian University ---
  Moscow Timiryazev Agricultural Academy, Moscow, 127550, Russia}
\end{center}
\vskip 2cm
\begin{abstract}
In the present paper we study equations of state of Casimir vacuum of massive scalar field and massive bispinor field in compact Friedmann Universe. With use of the Abel -- Plana formula the renormalization of divergent series for calculation of the quantum means of operators is implemented.
\end{abstract}

\newpage


\section{Introduction}

The Casimir effect plays an essential role in spaces with nontrivial topology and cosmological models~\cite{MTrunov}.
In studying the processes occurring in the early Universe, the influence of the quantum vacuum is significant~\cite{Mamaev:1976zb}. To obtain the components of the energy--momentum tensor of massive scalar field a regularized method of wavelength cutoff was utilized in Ref. 3.
The static Casimir condensate of conformal scalar field in Friedmann Universe was obtained in Ref. 4 by Abel--Plana regularized method.
The vacuum energy density of massive boson field is positive, whereas the one of massive fermion field is negative in Minkowski spacetime~\cite{Martin:2012bt}. It gives hope to solve the cosmological problem~\cite{Weinberg}.
It is of interest to consider the manifestation of the Casimir effect in a non-trivial space that is actual in cosmology especially in the early Universe. The equation of state of matter $p=w\epsilon$ connects the pressure $p$ and the energy density $\epsilon$, where $w$ is a proportionality coefficient: for interstellar dust $p=0$; for radiation $p=\epsilon/3$; for $\Lambda$-term contribution $p=-\epsilon$~\cite{RiessNobel}; for stiff state of matter $p=\epsilon$~\cite{Zakharov:2010nf, PavlovMIPh,Pervushin:2017zfj}.

In the present paper we study the pure Casimir components of the vacuum energy--momentum tensor, considering the quasi--static case. During the evolution of the Universe, the masses of elementary particles grow, which also manifests itself in the characteristics of the quantum vacuum. Quantum vacuum has unusual properties.
So, it is of interest to obtain the equation of state of vacuum for each moment of the Universe evolution. The Abel--Plana formula from the theory of analytical functions is applied for renormalization of ultraviolet divergences as the effective method. The difference between the divergent functional series and the corresponding integral in the mathematical Abel--Plana formula corresponds to the difference between the dynamical characteristics of the fields on the 3-sphere and its tangent flat space.
The natural system of units is used in calculations below $c=\hbar=G=1$.

\section{Quantum vacuum of massive scalar field}

\begin{figure}
\centering
{\vbox{
\psfig{figure=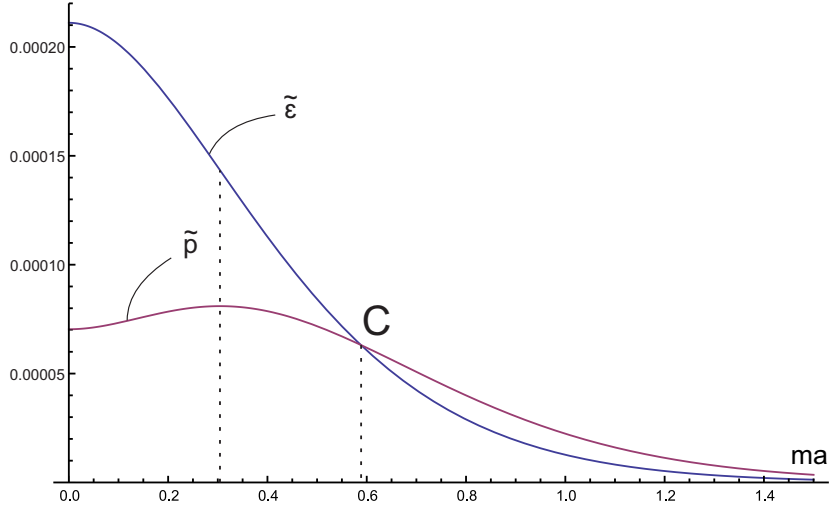, width=4.5in, bbllx=0pt, bblly=0pt,
bburx=420pt, bbury=250pt,
clip=}}}
\par
\caption{{Casimir energy density and pressure of the massive scalar field. Here we put $\tilde\epsilon=a^4\epsilon,$ $\tilde{p}=a^4p$
}}\label{Bose}
\end{figure}
The spacetime metric of a Friedmann universe
${\cal M}=\mathbb{R}^1\times S^3$ corresponds to the sphere of radius $a (\eta)$
\be \label{ds}
ds^2=a^2(\eta)\left(-d\eta^2+d\chi^2+\sin^2\chi(d\theta^2+\sin^2\theta d\phi^2)\right),
\ee
where $\eta$ is the conformal time; $a(\eta)$ is the conformal scale factor
having the sense of the $S^3$-sphere radius; $\chi$, $\theta$, and $\phi$ are
dimensionless angular coordinates on the sphere.

The quantum mean value of the time-time component of the energy--momentum tensor of the massive scalar field~\cite{MTrunov}
\be
\langle 0|T_{00}|0\rangle=\frac{1}{4\pi^2 a^2}\sum\limits_{\lambda=1}^\infty\lambda^2\sqrt{\lambda^2+m^2a^2}.
\ee
After application the Abel -- Plana formula we get the renormalized value for the energy density
(see Fig.~\ref{Bose})
\be\label{epsenergy}
\epsilon=
\frac{1}{2\pi^2a^4}\int\limits_{ma}^\infty
d\lambda\frac{\lambda^2\sqrt{\lambda^2-m^2a^2}}{\exp (2\pi\lambda)-1}.
\ee

The quantum mean values of the spatial components of the energy--momentum tensor of the massive scalar field: \cite{MTrunov}
\be
\langle 0|T_{ij}|0\rangle=
\frac{1}{12\pi^2 a^2}\sum\limits_{\lambda=1}^\infty\frac{\lambda^4}{\sqrt{\lambda^2+m^2a^2}}\gamma_{ij},
\ee
where $\gamma_{ij}$ are the spatial metric components.
After application of the Abel -- Plana formula we get the renormalized value for the pressure (see Fig.~\ref{Bose})
\be\label{epsp}
{p}=
\frac{1}{6\pi^2a^4}\int\limits_{ma}^\infty d\lambda
\frac{\lambda^4}{\sqrt{\lambda^2-m^2a^2}(\exp (2\pi\lambda)-1)}.
\ee

\begin{figure}
\centering
{\vbox{
\psfig{figure=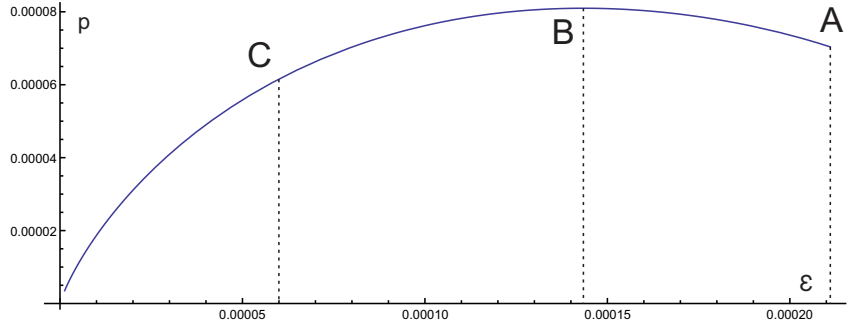, width=4.5in, bbllx=0pt, bblly=0pt,
bburx=500pt, bbury=250pt,
clip=}}}
\par
\caption{{Equation of state of the massive scalar field vacuum for every instant $ma$
}}\label{EoS}
\end{figure}
The set of functions $\epsilon = \epsilon(ma)$ (\ref{epsenergy}) and
${p} = {p}(ma)$ (\ref{epsp}) can be considered as the parametric representation  of the equation of state of the vacuum ${p}={p}(\epsilon)$ (see Fig.~\ref{Bose}). The explicit dependence $p=p(\epsilon)$ is shown in Fig.~\ref{EoS}.
In the massless limit $(ma=0)$ equation of state is ultrarelativistic (Fig.~\ref{EoS}, point $A$)
\be\nonumber
\epsilon=3p=\frac{1}{480\pi^2 a^4}.
\ee
In the region $ABC$ (Fig.~\ref{EoS}) the dominant energy condition takes place.
The vacuum of virtual particles in case of $p=\epsilon$ $(ma\approx 0.6)$ is in the rigid state (Fig.~\ref{Bose}, Fig.~\ref{EoS}, point C). Thus, we have EoS $p=w\epsilon$ with $1/3\leq w\leq 1$ in the region ABC (Fig.~\ref{EoS}).
With increasing the mass $ma$, the condition of energodominance is in violence for real but not for virtual particles
(to the right from point $C$ in Fig.~\ref{Bose}, and to the left from point $C$ in Fig.~\ref{EoS}).
In the limit $ma\gg 1$ the Casimir energy density as well as the pressure are exponentially small
\be\nonumber
\epsilon\approx\frac{(ma)^{5/2}}{8\pi^3a^4}e^{-2\pi ma},\qquad
p\approx\frac{(ma)^{7/2}}{12\pi^2a^4}e^{-2\pi ma}.
\ee

\section{Quantum vacuum of massive bispinor field}

The quantum mean value of the time-time component of the energy--momentum tensor of the massive bispinor field~\cite{MTrunov}:
\be
\langle 0|T_{00}|0\rangle=-\frac{1}{\pi^2}
\sum\limits_{\lambda=3/2}^\infty\left(\lambda^2-\frac{1}{4}\right)\sqrt{\lambda^2+m^2a^2}.
\ee
After application the Abel--Plana formula we get the energy density of the massive bispinor field~\cite{MTrunov}
\be\label{tildeepsilon}
\epsilon=\frac{2}{\pi^2a^4}\int\limits_{ma}^\infty\,d\lambda
\frac{(\lambda^2+1/4)\sqrt{\lambda^2-m^2a^2}}{\exp (2\pi\lambda)+1}.
\ee
\begin{figure}
\centering
{\vbox{
\psfig{figure=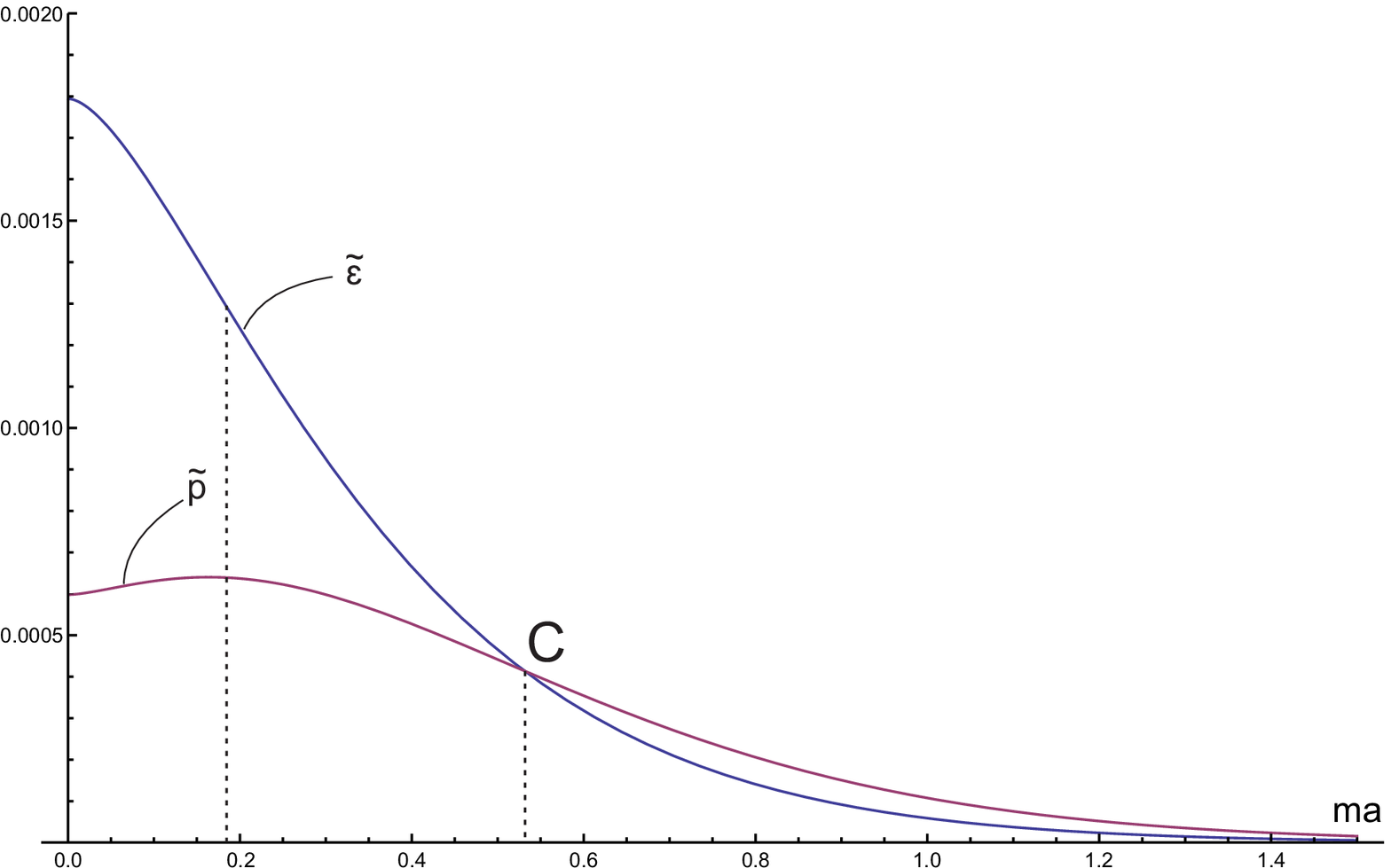, width=4.5in, bbllx=0pt, bblly=0pt,
bburx=520pt, bbury=300pt,
clip=}}}
\par
\caption{{Casimir energy density and pressure of the massive bispinor field. Here we put $\tilde\epsilon=a^4\epsilon,$ $\tilde{p}=a^4p$
}}\label{Fermion}
\end{figure}

\begin{figure}
\centering
{\vbox{
\psfig{figure=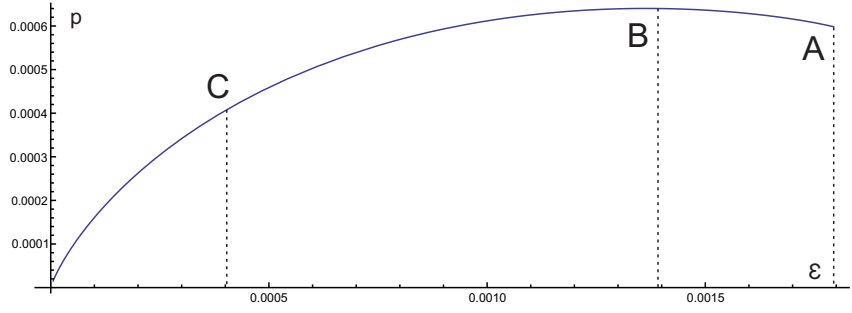, width=4.5in, bbllx=0pt, bblly=0pt,
bburx=520pt, bbury=250pt,
clip=}}}
\par
\caption{{Equation of state of the massive bispinor field vacuum for every instant $ma$
}}\label{EoSfermion}
\end{figure}

The vacuum expectation values of spatial components of the momentum-energy tensor are the following~\cite{MTrunov}:
\be
\langle 0|T_{ij}|0 \rangle=-\frac{1}{3\pi^2 a^2}
\sum\limits_{\lambda=3/2}^\infty\left(\lambda^2-\frac{1}{4}\right)\frac{\lambda^2}{\sqrt{\lambda^2+m^2a^2}}\gamma_{ij}.
\ee
Having used the Abel -- Plana formula, one obtains the renormalized value of the Casimir pressure
\be\label{tildep}
{p}=\frac{2}{3\pi^2 a^4}\int\limits_{ma}^\infty\,d\lambda
\frac{(\lambda^2+1/4)\lambda^2}{\sqrt{\lambda^2-m^2a^2}(\exp (2\pi\lambda)+1)}.
\ee
The corresponding figures of the functions $\epsilon =\epsilon (ma)$, ${p}={p}(ma)$ are shown in Fig.~\ref{Fermion}. The explicit dependence ${p}={p}(\epsilon)$ for every $ma$ as the equation of Casimir vacuum state is shown in Fig.~\ref{EoSfermion}.
In the massless limit $(ma=0)$ the equation of state is ultrarelativistic (Fig.~\ref{EoSfermion}, point $A$)
\be\nonumber
\epsilon=3p=\frac{17}{960\pi^2 a^4}.
\ee
In the region $ABC$ (Fig.~\ref{EoSfermion}) the dominant energy condition is true.
The vacuum of virtual fermion particles in case of $p=\epsilon$ $(ma\approx 0.55)$ is in the rigid state
(Fig.~\ref{Fermion}, Fig.~\ref{EoSfermion}, point C). The EoS $p=w\epsilon$ with $1/3\le w\le 1$ is presented in the region ABC in Fig.~\ref{EoSfermion}.
With increasing the mass $ma$, the condition of energodominance having sense for real particles is in violence
(to the right from point $C$ in Fig.~\ref{Fermion}, and to the left from point $C$ in Fig.~\ref{EoSfermion}).
In the limit $ma\gg 1$ the Casimir energy density as well as the pressure are exponentially small.

\section{Conclusion}

With use of the Abel -- Plana formula the renormalization of divergent series for calculation of characteristics of quantum vacuum is implemented. The topological Casimir effect manifests itself in Compton times of the Universe evolution.
It turns out that the Casimir energy density and pressure of both the boson and fermion fields is positive in Friedmann space. This is how the nontrivial topology of the space appears.
The equations of state $p=w\epsilon$ of Casimir vacuum of massive scalar field and bispinor massive field in compact Friedmann Universe is presented graphically.

\section{Acknowledgement}

I am grateful to Professor A. B. Arbuzov for fruitful discussions.


\end{document}